\documentclass[10pt,amsmath,amssymb,nofootinbib,aps,prl]{revtex4}
\begin{document}

\title{Neutrino superluminality without Cherenkov-like processes in Finslerian special relativity}

\author{Zhe Chang}
\email{changz@ihep.ac.cn}
\author{Xin Li}
\email{lixin@ihep.ac.cn}
\author{Sai Wang}
\email{wangsai@ihep.ac.cn}

\affiliation{Institute of High Energy Physics\\
Theoretical Physics Center for Science Facilities\\
Chinese Academy of Sciences\\
100049 Beijing, China}

\begin{abstract}

Recently, Cohen and Glashow (Phys. Rev. Lett. {\bf 107}, 181803 (2011)) pointed out that the superluminal neutrinos reported by the OPERA
would lose their energy rapidly via the Cherenkov-like process.
The Cherenkov-like process for the superluminal particles would be forbidden
if the principle of special relativity holds in any frame instead violated with a preferred frame.
We have proposed that the Finslerian special relativity could account for the data of the neutrino superluminality (arXiv:1110.6673[hep-ph]).
The Finslerian special relativity preserves the principle of special relativity and involves a preferred direction while consists with the causality.
In this paper, we prove that the energy-momentum conservation is preserved and the energy-momentum is well defined in Finslerian special relativity.
The Cherenkov-like process is forbidden in the Finslerian special relativity.
Thus, the superluminal neutrinos would not lose energy in their distant propagation.

\end{abstract}

\maketitle

\section{I. Introduction}

Recent report by the OPERA Collaboration showed possibility that the muon neutrinos could be superluminal \cite{OPERA2011}.
The previous experiments or observations \cite{MINOS2007,Fermilab1979,SN1987A} also gave constraints on the neutrino superluminality.
In this paper, the natural units are used which imply \(c=1\).
The superluminality of particles is forbidden in Einstein's special relativity.
The speed of light is the upper limit of speed for all particles in Einstein's special relativity,
unless the Lorentz invariance violation (LIV) \cite{LIVCG,LIVCK00,LIVCK,VSR} is involved.
Soon after release of the OPERA's report, Cohen and Glashow \cite{CohenG01} pointed out that
the number of superluminal neutrinos with energy above \(12.5~\rm{GeV}\) should be strongly suppressed
after a distant propagation from CERN to Gran Sasso laboratory.
The superluminal neutrinos would lose their energy rapidly via the Cherenkov-like process (\(\nu\longrightarrow\nu+e^{-}+e^{+}\))
in the context of LIV with a preferred frame.
In this preferred frame only, the energy-momentum conservation is preserved \cite{VillanteV001,Amelino-CameliaETAL01}.
Bi {\it et al.} \cite{BiETAL01} made a similar discussion on this issue.
Furthermore, Li {\it et al.} \cite{LiLMWZ001} pointed that the Cherenkov-like process even exists
in the trivial frame without the effective rest frame.
Nevertheless, Amelino-Camelia {\it et al.} \cite{Amelino-CameliaETAL01} recently showed that the Cherenkov-like process may be forbidden
in a context that the principle of special relativity is preserved in any frames while the energy-momentum conservation is modified.

If the neutrino superluminality is confirmed, Einstein's special relativity as well as the Minkowski description of spacetime should be amended.
In the new spacetime, the causality still holds and the superluminality is admitted without the Cherenkov-like process.
In our previous work \cite{Finsler special relativity}, an alternative special relativity in the Finsler spacetime \cite{Book by Bao,Book by Shen},
so-called Finslerian special relativity, was proposed to account for the reported neutrino superluminality.
In Finslerian special relativity, a preferred direction is involved in the spacetime background, which introduces the superluminality of particles.
The superluminality in Finslerian special relativity was found to be linearly dependent on the energy per unit mass of the particles.
It was revealed that Finslerian special relativity is roughly consistent with the data of the neutrino superluminality
from the present experiments and observations.
In addition, the superluminality of particles is not contradictory with the causality.

The Finsler spacetime is a straightforward generalization of the Riemann spacetime.
It abandons the restriction on the quadratic form of the metric, which may lead to new insights on the spacetime structure.
The Finsler spacetime structure is dependent on one or more preferred directions.
The LIV has been studied in the Finsler spacetime with modified dispersion relation (MDR) \cite{Finsler special relativity,MDR,GirelliLS001}.
In addition, the very special relativity (VSR) \cite{CohenG001} was proved to be a kind of Finslerian special relativity \cite{VSR Finsler}.
The other example of Finslerian special relativity \cite{ChangL001} resides in the Randers spacetime \cite{Randers space}.
Furthermore, the symmetry of Finslerian special relativity with constant curvature was studied in detail \cite{LiC001}.
The Finsler geometry has also brought out new insights on resolution of anomalies in Einstein's gravity and cosmology \cite{Finsler gravity}.

In this paper, we prove that the conservation of energy and momentum of particles is preserved in Finslerian special relativity.
The Finslerian line element is invariant under the spacetime translations.
Combining with the energy-momentum conservation and the MDR in Finslerian special relativity,
we show that the Cherenkov-like process is forbidden for the superluminality of neutrinos.
The superluminal neutrinos would not lose energy in their distant propagation.
Thus, Finslerian special relativity admits the existence of the superluminal neutrinos as well as their distant propagation.
The rest of the paper is arranged as follows.
In Section II, main points of Finslerian special relativity is summarized briefly and the superluminality of particles is showed in this picture.
In Section III, we demonstrate that the energy-momentum conservation still hold in Finslerian special relativity
and there exists no Cherenkov-like process for the superluminal particles.
The conclusions and remarks are given in Section IV.

\section{II. Superluminality in Finslerian special relativity}

Based upon the reports on the superluminal neutrinos,
we have proposed Finslerian special relativity \cite{Finsler special relativity} to account for the observations.
As mentioned in the Introduction, Finslerian special relativity admits the neutrino superluminality and preserves the causality.
Finslerian special relativity involves a preferred direction which leads to the superluminality of particles.
In Finslerian special relativity, the superluminal behaviors of particles are found to be linearly dependent on their energy per unit mass,
which is roughly consistent with the data of the reports from the present neutrino experiments.
In this section, we summarize the outlines of Finslerian special relativity and reveal its superluminality.

The action of free particles in Finslerian special relativity is of the form
\begin{equation}
\label{integral length}
I\propto\int^b_a F\left(x, y\right)d\tau\ ,
\end{equation}
where $x^{\mu}$ denotes the position and \(y^{\mu}:=dx^{\mu}/d\tau\) denotes the 4-velocity of particles.
Note that the Greek indices run from $0$ to $3$ and the Latin indices run from $1$ to $3$ in this Letter.
The Finsler structure \(F\) is positively homogeneous of order one.
To account for the data of the neutrino superluminality, Finsler line element
of the (\(\alpha,\beta\)) type \cite{alpha beta type} was proposed as \cite{Finsler special relativity}
\begin{equation}
\label{Finsler Superluminal}
F(y)d\tau=\alpha\left(1-A\left(\frac{{\beta}}{{\alpha}}\right)^{3}\right)d\tau\ ,
\end{equation}
where
\begin{eqnarray}
\alpha&=&\sqrt{\eta_{\mu\nu}y^{\mu}y^{\nu}}\ ,\\
\beta&=&b_{\mu}y^{\mu}\ .
\end{eqnarray}
Here, \(\eta_{\mu\nu}\) denotes the Minkowski metric \((+1,-1,-1,-1)\), $b_{\mu}$ denotes constant vector \((1,0,0,0)\)
and $A$ denotes a tiny positive dimensionless constant (\(A\ll 1\))
which is determined uniquely by the data of the neutrino superluminality.
The timelike vectors, such as \(b_{\mu}\), are well defined in Finsler spacetime (\ref{Finsler Superluminal}),
with the usual property of forming a convex cone in each tangent space \(T_{x}M\). They satisfy the condition that \(g^{\mu\nu}b_{\mu}b_{\nu}>0\), where the fundamental tensor \(g_{\mu\nu}\) is defined as \(g_{\mu\nu}:=\frac{1}{2}\frac{\partial^{2}F^{2}}{\partial y^{\mu}\partial y^{\nu}}\).
In the flat Finsler spacetime, the tangent spaces at each point are isomorphic \cite{Book by Bao}.
Thus, it is possible to define well the timelike vectors at each tangent space \(T_{x}M\) in the spacetime (\ref{Finsler Superluminal}).
This flat Finslerian line element departs mildly from the Minkowski line element and it returns back to the Minkowski one in the case of \(A=0\).
Thus, Finslerian special relativity is a mild generalization of Einstein's special relativity.

As is defined in Einstein's special relativity, the canonical 4-momentum is defined as \cite{GirelliLS001}
\begin{equation}
p_{\mu}:=m\frac{\partial F}{\partial y^{\mu}}\ .
\end{equation}
The 4-velocity is defined as
\begin{equation}
\label{velocity in null structure}
u_{\mu}:=\frac{\partial F}{\partial y^{\mu}}\ ,
\end{equation}
which is the canonical momentum divided by mass.
The null structure in Finslerian special relativity is given by \(F(y)=0\).
The null structure is revealed finally by
\begin{equation}
\eta^{\mu\nu}u^{'}_{\mu}u^{'}_{\nu}+2A(u^{'}_{0})^{3}=0\ ,
\end{equation}
where we have neglected the terms of higher orders in $A$ and the primes denote the normalization with $F$.
Then the causal speed is given by
\begin{equation}
v_{c}:=\frac{\sqrt{-\eta^{ij}u^{'}_{i}u^{'}_{j}}}{\sqrt{\eta^{00}u^{'}_{0}u^{'}_{0}}}\approx1+Au^{'}_{0}\ .
\end{equation}
It is noted that the causal speed is enlarged in Finslerian special relativity than that in Einstein's special relativity.

For the particles with mass, the normalization of the Finsler norms is realized by \(F(y)=1\).
The canonical 4-momentum of the particle with mass \(m\) is given by
\begin{equation}
\label{4-momentum}
p_{\mu}=m\frac{\partial F}{\partial y^{\mu}}\ .
\end{equation}
The physical dispersion relation corresponding to Finslerian special relativity is given as
\begin{equation}
\label{MDR}
\eta^{\mu\nu}p_{\mu}p_{\nu}+\frac{2A}{m}\left(p_{0}\right)^{3}=m^{2}\ .
\end{equation}
Then the speed of the particle is obtained as
\begin{equation}
\label{superluminal speed formula}
v:=\frac{\sqrt{-\eta^{ij}p_{i}p_{j}}}{\sqrt{\eta^{00}p_{0}p_{0}}}=1-\frac{1}{2u^{2}}+Au\ ,
\end{equation}
where \(u\) denotes the energy per unit mass \({E}/{m}\) of the particle.
In the case of \(u\) large enough, the superluminal behaviors of particles emerge.
By comparing the speed formula (\ref{superluminal speed formula}) with the data of the present neutrino superluminality,
we constrain the parameter \(A\) to be of order \(10^{-18}\) \cite{Finsler special relativity}.

\section{III. Finslerian special relativity admits NO Cherenkov-like process}

The superluminality is stringently forbidden in Einstein's special relativity
and the superluminal particles would lose their energy rapidly via the Cherenkov-like process even in the context of LIV with a preferred frame.
Amelino-Camelia {\it et al.} argued that the context, in which the principle of special relativity still holds
while the spacetime structure is influenced by the quantum gravity, may forbid the Cherenkov-like process.
The Finslerian special relativity mildly departs from Einstein's special relativity
since the spacetime structure in Finslerian special relativity mildly deviates from the Minkowski one.
Finslerian special relativity may also forbid the Cherenkov-like process.

It is obvious that Finslerian special relativity is invariant under the spacetime translations,
since the line element (\ref{Finsler Superluminal}) of Finslerian special relativity contains no dependence on the spacetime positions.
This could also be demonstrated by the Killing vectors approach \cite{LiC001}.
The infinitesimal coordinate transformation is
\begin{eqnarray}
x^{\mu}&\longrightarrow& x^{\mu}+\epsilon V^{\mu}\ ,\\
y^{\mu}&\longrightarrow& y^{\mu}+\epsilon \frac{\partial V^{\mu}}{\partial x^{\nu}}y^{\nu}\ ,
\end{eqnarray}
where \(|\epsilon|\ll 1\) and the generators are called Killing vectors \(V^{\mu}\).
Under the above coordinate transformation, a Finsler structure is called isometry if and only if
\begin{equation}
F(x,y)=F(\bar{x},\bar{y})\ .
\end{equation}
For the Finslerian line element (\ref{Finsler Superluminal}) of (\(\alpha,\beta\)) type,
the isometric transformation implies that the Killing vectors satisfy the Killing equations
\begin{equation}
\label{Killing equations}
V^{\mu}\frac{\partial F}{\partial x^{\mu}}+y^{\nu}\frac{\partial V^{\mu}}{\partial x^{\nu}}\frac{\partial F}{\partial y^{\mu}}=0\ .
\end{equation}
It is obvious that the constant vectors \(C^{\mu}\) are solutions of the above Killing equations
for the Finslerian line element (\ref{Finsler Superluminal}).
Based on the Noether theorem, the spacetime translational invariance implies that
the energy-momentum \(p_{\mu}\) is well defined and conserved in Finslerian special relativity.

In the following, we prove that the energy-momentum conservation and the MDR in Finslerian special relativity are enough
to reveal that the superluminal neutrinos would not lose their energy via the Cherenkov-like process.
The Cherenkov-like process is forbidden in Finslerian special relativity.
After a distant propagation,
a large number of superluminal neutrinos survive and could be received by the OPERA detector.
Thus, the stringent constraint proposed by Cohen and Glashow on the superluminality
would not play part in the superluminality of neutrinos in Finslerian special relativity.
It is enough to describe properties of the Cherenkov-like process through the process \(\mu\longrightarrow M+M\).
There are one single incoming particle with mass \(\mu\), energy \(E\), and momentum \({P}\)
while two ejected particles with mass both \(M\), energy \(E_{1}\), \(E_{2}\),
and momentum \({P}_{1}\), \({P}_{2}\) \cite{Amelino-CameliaETAL01}.
Meanwhile, the two ejected particles are heavier than the incoming particle, namely \(\mu<M\).
In Finslerian special relativity, the energy and momentum conservations imply that
\begin{eqnarray}
E&=&E_{1}+E_{2}\ ,\\
{P}^{2}&=&{P}_{1}^{2}+{P}_{2}^{2}+2P_{1}P_{2}\cos\theta\ ,
\end{eqnarray}
where \(\theta\) denotes the angle between the moving directions of the two ejected particles.
By combining the definition of the 4-momentum (\ref{4-momentum}) and the MDR (\ref{MDR}) with the energy-momentum conservation relations,
we obtain
\begin{eqnarray}
\cos{\theta}&=&\frac{2E_{1}E_{2}+2A\frac{M-\mu}{\mu M}(E_{1}^{3}+E_{2}^{3})+
\frac{6A}{\mu}E_{1}E_{2}(E_{1}+E_{2})-\mu^{2}+2M^{2}}{2E_{1}E_{2}+\frac{2A}{M}E_{1}E_{2}(E_{1}+E_{2})-
M^{2}\left(\frac{E_{1}}{E_{2}}+\frac{E_{2}}{E_{1}}\right)}+\mathcal{O}(A^{2})\nonumber\\
&=&1+\frac{2A\frac{M-\mu}{\mu M}(E_{1}^{3}+E_{2}^{3})+
\frac{4A}{\mu}E_{1}E_{2}(E_{1}+E_{2})-\mu^{2}+2M^{2}+M^{2}\left(\frac{E_{1}}{E_{2}}+\frac{E_{2}}{E_{1}}\right)}
{2E_{1}E_{2}+\frac{2A}{M}E_{1}E_{2}(E_{1}+E_{2})-
M^{2}\left(\frac{E_{1}}{E_{2}}+\frac{E_{2}}{E_{1}}\right)}+\mathcal{O}(A^{2})\nonumber\\
&>&1+\frac{2A\frac{M-\mu}{\mu M}(E_{1}^{3}+E_{2}^{3})+
\frac{4A}{\mu}E_{1}E_{2}(E_{1}+E_{2})}{2E_{1}E_{2}+\frac{2A}{M}E_{1}E_{2}(E_{1}+E_{2})}+\mathcal{O}(A^{2})\ ,
\end{eqnarray}
where the ultra relativistic approximation is involved (\(\mu\ll E,~M\ll E_{1},~M\ll E_{2}\)) in the last step.

The mass of incoming particle is smaller than that of outcoming particles, namely \(\mu<M\).
The OPERA experiment of superluminal neutrinos indicates that the Finsler parameter \(A\) has a positive tiny value \(\mathcal{O}(10^{-18})\) \cite{Finsler special relativity}.
Therefore, the fraction of the second term is positive in the final expression of Eq.(18).
Thus, we find that \(\cos\theta\) is always greater than \(1\) in Eq.(18)
which reveals that the Cherenkov-like process is forbidden in Finslerian special relativity.
Thus, the Cohen-Glashow constraint on the superluminality does not impact on
the superluminal behaviors of neutrinos in Finslerian special relativity.
The constraint on the superluminality proposed by Cohen and Glashow showed that
the OPERA's report could not be explained by the propagation of the superluminal neutrinos in the framework of LIV with the preferred frame.
We showed that this constraint does not exist in Finslerian special relativity.
Thus, the OPERA's result could be interpreted with the viewpoint of the propagation of superluminal neutrinos
in the framework of Finslerian spacial relativity.

\section{IV. Conclusions and remarks}

If the superluminal behaviors of particles are confirmed, they would destroy the theoretical system of the standard theories.
Even in the context of LIV with preferred frame, the superluminality of particles is suppressed strongly via the Cherenkov-like process.
However, Amelino-Camelia {\it et al.} argued that the superluminality is admitted
in the framework which still preserves the principle of relativity and the energy-momentum conservation.
Finslerian special relativity could be a reasonable candidate to realize this destination.

Finslerian special relativity is a mild generalization of Einstein's special relativity.
The Finslerian line element departs with a preferred direction from the Minkowski one.
The dispersion relation in Finslerian special relativity deviates from the one in Einstein's special relativity, which is corresponding to the superluminality of particles.
In Finslerian special relativity, the superluminal behaviors of particles are permitted and the causal speed is enlarged but the causality still holds.
In addition, Finslerian special relativity could account for the data of the present superluminal neutrino experiments and observations.
In this paper, we showed that the energy and momentum of particles are conserved in Finslerian special relativity.
Then we proved that the Cherenkov-like process is forbidden in Finslerian special relativity.
The superluminal neutrinos would not lose energy in their distant propagations.

It is worthwhile to note that the Cherenkov-like process is analyzed as a quantum field theoretic (QFT) process.
We have not studied the QFT characters in Finsler spacetime.
However, it still makes sense to employ properties like the energy-momentum conservation and the MDR (\ref{MDR}).
The reason is that Finsler spacetime (\ref{Finsler Superluminal}) may be viewed as a realization
of the spontaneous breaking of Lorentz symmetry \cite{LIVCK00}, in which the energy-momentum conservation and MDR are employed.
The properties like the momentum conservation and MDR may still provide inspirations on the Cherenkov-like process.
Regardless the loop corrections, the kinematical results at the tree level would give meaningful implications on the Cherenkov-like process.
The spontaneous breaking of Lorentz symmetry means that there are certain fixed background fields
which determine the preferred directions of the spacetime.
Finsler geometry has natural advantage to deal with the preferred directions,
since Finsler metric tensors are intrinsically dependent on the directions of motions of particles \cite{Book by Bao,Book by Shen}.
Recently, Kostelecky \cite{KosteleckyFinsler} proposed that the classical Lagrangian with the spontaneous LIV corresponds
to one classic line element in Finsler spacetime.
In the spontaneous LIV models, the energy and momentum are conserved and the dispersion relation is modified.
The energy-momentum conservation and the MDR are assumed to deal with the perturbative QFT processes
in the standard model extension (SME) \cite{LIVCK} which is a spontaneous LIV model.
Finsler line element (\ref{Finsler Superluminal}) also depends on a preferred direction \(b_{\mu}\).
Thus, it could be viewed as an implication of the spontaneous breaking of Lorentz symmetry,
in which the LIV was dealt in the perturbative view of QFT in Minkowski spacetime.
Therefore, the classic properties may still provide inspirations on the Cherenkov-like process in Finsler spacetime.\\

\begin{acknowledgments}
We thank useful discussions with Y. G. Jiang, M. H. Li and H. N. Lin.
This work is supported by the National Natural Science Fund of China under Grant No. 10875129 and No. 11075166.

\end{acknowledgments}


\begin{thebibliography}{999}
%1
\bibitem{OPERA2011}T. Adam {\it et al.} [OPERA Collaboration], arXiv:1109.4897.
\bibitem{MINOS2007}P. Adamson {\it et al.} [MINOS Collaboration], Phys. Rev. D {\bf 76}, 072005 (2007).
\bibitem{Fermilab1979}J. Alspector {\it et al.}, Phys. Rev. Lett. {\bf 36}, 837 (1976).
    G. R. Kalbfleisch, N. Baggett, E. C. Fowler, and J. Alspector, Phys. Rev. Lett. {\bf 43}, 1361 (1979).
\bibitem{SN1987A}K. Hirata {\it et al.}, Phys. Rev. Lett. {\bf 58}, 1490 (1987).
    R. M. Bionta {\it et al.}, Phys. Rev. Lett. {\bf 58}, 1494 (1987).
    M. J. Longo, Phys. Rev. D {\bf 36}, 3276 (1987).
\bibitem{LIVCG}S. R. Coleman and S. L. Glashow, Phys. Lett. B {\bf 405}, 249 (1997).
    S. R. Coleman and S. L. Glashow, Phys. Rev. D {\bf 59}, 116008 (1999).
\bibitem{LIVCK00}V. A. Kostelecky and S. Samuel, Phys. Rev. D {\bf 39}, 683 (1989).
\bibitem{LIVCK}D. Colladay and V. A. Kostelecky, Phys. Rev. D {\bf 55}, 6760 (1997).
    D. Colladay and V. A. Kostelecky, Phys. Rev. D {\bf 58}, 116002 (1998).
%2
\bibitem{VSR}A. G. Cohen and S. L. Glashow, Phys. Rev. Lett. {\bf 97}, 021601 (2006).
\bibitem{CohenG01}A. G. Cohen and S. L. Glashow, Phys. Rev. Lett. {\bf 107}, 181803 (2011); arXiv:1109.6562 [hep-ph].
\bibitem{VillanteV001}F. L. Villante and F. Vissani, arXiv:1110.4591 [hep-ph].
\bibitem{Amelino-CameliaETAL01}G. Amelino-Camelia, L. Freidel, J. Kowalski-Glikman and L. Smolin, arXiv:1110.0521 [hep-ph].
\bibitem{BiETAL01}X. J. Bi, P. F. Yin, Z. H. Yu and Q. Yuan, Phys. Rev. Lett. {\bf 107}, 241802 (2011); arXiv:1109.6667 [hep-ph].
%3
\bibitem{LiLMWZ001}M. Li, D. Liu, J. Meng, T. Wang and L. Zhou, arXiv:1111.3294 [hep-ph].
\bibitem{Finsler special relativity}Z. Chang, X. Li and S. Wang, arXiv:1110.6673 [hep-ph], acceptted for publication by Mod. Phys. Lett. A.
\bibitem{Book by Bao}D. Bao, S. S. Chern, and Z. Shen, {\it An Introduction to Riemann--Finsler Geometry},
        Graduate Texts in Mathmatics {\bf 200}, Springer, New York, 2000.
\bibitem{Book by Shen}Z. Shen, {\it Lectures on Finsler Geometry}, World Scientific, Singapore, 2001.
\bibitem{MDR}D. Ratzel, S. Rivera, F. P. Schuller, Phys. Rev. D {\bf 83}, 044047 (2011); arXiv:1010.1369 [hep-th].
    G. Amelino-Camelia, J. R. Ellis, N. E. Mavromatos and D. V. Nanopoulos, Int. J. Mod. Phys. A {\bf 12}, 607 (1997); arXiv:hep-th/9605211.
    R. Gambini, J. Pullin, Phys. Rev. D {\bf 59}, 124021 (1999); arXiv:gr-qc/9809038.
    F. P. Schuller, C. Witte and M. N. R. Wohlfarth, Annals Phys. {\bf 325}, 1853 (2010); arXiv:0908.1016 [hep-th].
    C. Pfeifer and M. N. R. Wohlfarth, Phys. Rev. D {\bf 84} 044039 (2011); arXiv:1104.1079 [gr-qc].
    C. Pfeifer and M. N. R. Wohlfarth, arXiv:1109.6005 [gr-qc].
    A. Kostelecky, Phys. Lett. B {\it 701}, 137 (2011); arXiv:1104.5488 [hep-th].
\bibitem{GirelliLS001}F. Girelli, S. Liberati and L. Sindoni, Phys. Rev. D {\bf 75}, 064015 (2007); arXiv:gr-qc/0611024.
%4
\bibitem{CohenG001}A. G. Cohen, S. L. Glashow, Phys. Rev. Lett. {\bf 97}, 021601 (2006); arXiv:hep-ph/0601236.
\bibitem{VSR Finsler}G. W. Gibbons, J. Gomis, C. N. Pope, Phys. Rev. D {\bf 76}, 081701 (2007).
    G. Y. Bogoslovsky, arXiv:0706.2621.
    H. F. Goenner and G. Y. Bogoslovsky, Gen. Rel. Grav. {\bf 31}, 1383 (1999); arXiv:gr-qc/9701067.
    A. P. Kouretsis, M. Stathakopouslos, and P. C. Stavrinos, Phys. Rev. D {\bf 79}, 104011 (2009); arXiv:0810.3267.

\bibitem{ChangL001}Z. Chang and X. Li, Chinese Phys. C {\bf 33}, 626 (2009). Phy. Lett. B {\bf 663}, 103 (2008).
\bibitem{Randers space}G. Randers, Phys, Rev. {\bf 59}, 195(1941).
\bibitem{LiC001}X. Li and Z. Chang, arXiv:1010.2020 [gr-qc].
%5
\bibitem{Finsler gravity}
    Y. Takano, Lett. Nuovo Cimento {\bf 10}, 747 (1974).
    S. Ikeda, Ann. der Phys. {\bf 44}, 558 (1987).
    R. Tavakol, and N. van den Bergh, Phys. Lett. A {\bf 112}, 23 (1985).
    G. Yu. Bogoslovsky, Phys. Part. Nucl. {\bf 24}, 354 (1993).
    Z. Chang and X. Li, Phys.Lett.B {\bf 668}, 453 (2008).
    M. Milgrom, Astrophys. J. {\bf 270}, 365 (1983).
    Z. Chang and X. Li, Phys. Lett. B {\bf 676}, 173 (2009).
    X. Li, Z. Chang and M. H. Li, arXiv:1001.0066.
    Z. Chang, M. H. Li and X. Li, arXiv:1009.1509.
    X. Li and Z. Chang, Phys. Lett. B {\bf 692}, 1 (2010).
    J. D. Anderson {\it et al.}, Phys. Rev. Lett. {\bf 81}, 2858 (1998).
    J. D. Anderson {\it et al.}, Phys. Rev. Lett. {\bf 65}, 082004 (2002).
    J. D. Anderson {\it et al.}, Mod. Phys. Lett. A {\bf 17}, 875 (2002).
    X. Li and Z. Chang, Phys. Rev. D {\bf 82}, 124009 (2010).
    D. Clowe, S. W. Randall, and M. Markevitch, http://flamingos.astro.ufl.edu/1e0657/index.html; Nucl. Phys. B, Proc. suppl. {\bf 173}, 28 (2007).
    X. Li and Z. Chang, arXiv:0911.1890.
    Z. K. Silagadze, Acta Phys. Polon. B {\bf42}, 1199 (2011); arXiv:1007.4632.
    Z. Chang, S. Wang and X. Li, Eur. Phys. J. C {\bf 72}, 1838 (2012), DOI: 10.1140/epjc/s10052-011-1838-4; arXiv:1106.2726.
    X. Li and Z. Chang, arXiv:1108.3443.
    Z. Chang, M. H. Li, X. Li and S. Wang, arXiv:1110.3893 [astro-ph.CO].
\bibitem{alpha beta type}Z. Shen, {\it Some perspectives in Finsler geometry}, MSRI Publication Series. Cambridge: Cambridge university press, 2004.
\bibitem{KosteleckyFinsler}V. A. Kostelecky, Phys. Lett. B {\bf 701}, 137 (2011).











\end{thebibliography}
\end{document}